\documentclass[runningheads]{llncs}
\usepackage[utf8]{inputenc} 
\usepackage[T1]{fontenc}    
\usepackage{comment}

\usepackage{mathtools}
\usepackage{todonotes}
\usepackage{enumitem}
\usepackage{hyperref}
\usepackage{cleveref}
\usepackage{amsfonts}

\spnewtheorem{observation}{Observation}{\bfseries}{\normalfont}

\AtEndEnvironment{proof}{\phantom{}\qed}
\crefname{observation}{Observation}{Observations}
\crefname{property}{Property}{Properties}

\newcommand{\Overhead}{\mathcal{O}((\Delta~-~1)^{\delta} ( \Delta \log(n) + s(n) + \ell(n)))}
\newcommand{\LbProp}{\mathcal{P}_\Delta}

\newcommand{\packet}[1]{\texttt{pt}(#1)}

\newcommand{\Spacket}[1]{\texttt{pts}(#1)}
\newcommand{\bitstrings}{\{0, 1\}^*}
\newcommand{\PacExp}{(D, L, C, \omega, d)}
\newcommand{\Orig}[1]{\omega(#1)}

\newcommand{\A}{\mathcal{A}}
\newcommand{\Adefault}{\mathcal{A}(v)}
\newcommand{\Prop}{\mathcal{P}}
\newcommand{\bigO}{\mathcal{O}}
\newcommand{\has}[1]{\texttt{has-pt}(#1)}
\newcommand{\Nr}[2]{V[#1,#2]}

\newcommand{\LBvalBig}{\frac{n\cdot\varepsilon }{12 (\Delta - 1)^{r}}}
\newcommand{\LBval}{(n\cdot \varepsilon)/(12 (\Delta - 1)^{r})}
\newcommand{\UBval}{C \cdot n / (\Delta-1)^{r - 1}}

\newcommand{\visPackets}[1]{\mathcal{B}[#1]}

\begin{document}

\title{Decreasing verification radius in local certification
\thanks{The paper is eligible for the best student paper award.}
    } 

 \author{Laurent Feuilloley\inst{2}\orcidID{0000-0002-3994-0898}
 \and Jan Janoušek \inst{1}\orcidID{0000-0001-9329-1000}
 \and Jan Matyáš Křišťan\inst{1,3}\orcidID{0000-0001-6657-0020} \and
 Josef Erik Sedláček\inst{1,4}\orcidID{0009-0001-7429-2937}}


\authorrunning{L. Feuilloley, J. Janoušek, J.\,M. Křišťan and J.\,E. Sedláček} 



 \institute{Faculty of Information Technology, CTU in Prague, Czech Republic
 \and{CNRS, INSA Lyon, UCBL, LIRIS, UMR5205, F-69622 Villeurbanne, France}
 \and
 \email{kristja6@fit.cvut.cz}
 \and \email{sedlajo5@fit.cvut.cz}
 }

\newenvironment{retheorem}[1]{%
  \renewcommand\thetheorem{\ref{#1}}%
  \theorem%
}{\endtheorem}

\maketitle

\begin{abstract}
    This paper deals with \emph{local certification}, specifically locally checkable proofs: given a \emph{graph property}, the task is to certify whether a graph satisfies the property. The verification of this certification needs to be done \emph{locally} without the knowledge of the whole graph.
    More precisely, a distributed algorithm, called a \emph{verifier}, is executed on each vertex. The verifier observes the local neighborhood up to a constant distance and either accepts or rejects.

    We examine the trade-off between the visibility radius and the size of certificates. We describe a procedure that decreases the radius by encoding the neighbourhood of each vertex into its certificate.
    We also provide a corresponding lower bound on the required certificate size increase, showing that such an approach is close to optimal.


 \end{abstract}
    

\section{Introduction}
The problem studied in this paper involves certifying a global graph property without having knowledge of the entire graph.
In particular, we study the model of locally checkable proofs of Göös and Suomela~\cite{goos2016locally}.


In this model, a distributed algorithm called a \emph{verifier} examines the local neighbourhood of each vertex up to some fixed distance, called the \emph{radius}.
On each vertex, the verifier either accepts if it cannot deny that the graph has the desired property, or rejects if it is certain that the property is not satisfied. 
The final decision about the property is then made as follows:
If the verifier rejected on at least one vertex, the decision is that the property is not satisfied. If it accepts on all vertices, the decision is that the property holds.


To enhance the decision-making capabilities of the model, the vertices are equipped with unique identifiers and possibly more general labels.
Furthermore, each vertex is given a \emph{certificate}.

Certificates, are bit-strings that are used to help the verifier in deciding the answer about the property. 
The verifier reads the certificates in its local view as a part of its input.
For each graph that satisfies the property, the verifier must accept for at least one assignment of certificates. 
If the graph does not satisfy the property, the verifier must reject every assignment of certificates.


The key notion of local certification is that of a \emph{proof labeling scheme}, which is a pair $(f, \mathcal{A})$, where $\mathcal{A}$ is the verifier and $f$, often called a \emph{prover}, gives each graph with the property an assignment of certificates that is accepted by $\mathcal{A}$.

An intuitive example of a problem requiring certificates is $k$-colorability, where $k$ is a constant.
Clearly, $\mathcal{O}(\log k)$ bits are enough, as the coloring can be provided in the certificates.
If the graph is not $k$-colorable, then at least one vertex will be adjacent to a vertex with the same color (or use a color greater than $k$) and it would reject.




\subsection{Previous work}
Similar models, related to the LOCAL model \cite{peleg2000distributed}, have been studied under different names \cite{KormanKP10,goos2016locally}.
The name \emph{local certification} is a general term used for the similar models~\cite{feuilloley2021introduction}.

The concept of local certification is relevant to self-stabilization \cite{Dolev2000,AltisenDDP19}, as it is a component of many self-stabilizing algorithms \cite{KormanKP10}.
 

Since local certification has been introduced, lower bounds and upper bounds for the size of the certificates for many graph properties and problems have been proven \cite{censor2020approximate,ardevol2022lower,FEUILLOLEY20239,kormanKuttenSpanningTrees}. Also, the strength of the model in general and under various restrictions was studied \cite{fraigniaud2013towards,fraigniaud2018node,fraigniaud2013can}. For a general survey see~\cite{feuilloley2021introduction}.

It has been previously shown how, and under which conditions, certificate size can be decreased at the cost of increasing the visibility radius~\cite{Ostrovsky2017,FeuilloleyFHPP21,fischer2022explicit}.
We provide a similar result, showing how the visibility radius can be decreased at the cost of increasing the certificate size.

There is a subtle but crucial distinction between these two problems.
While the previous results allow increasing the radius while decreasing the size of certificates in the general case, the implied inverse procedure of decreasing the radius and increasing the certificate size works only for the very particular type of proof labeling schemes that result from the original procedure.
The novelty of our results lies in allowing the decrease of the radius of \emph{any} proof labeling schemes.

\subsection{Our contribution}
We show a general procedure for decreasing the radius of a proof labeling scheme at the expense of increasing the size of its certificates.
We also provide a corresponding lower bound on the necessary certificate size increase.

Consider the following motivation for this result.
Given a network, it may be needed to check that it satisfies a given property, such as absence of a cycle.
Communication between nodes of the network may be bounded by distance, which corresponds to the visibility radius of the verifier.
Under these conditions, proof labeling schemes provide a template for verifying properties of this network.
Our procedure then allows to decrease the communication distance at the cost of increasing the memory and computational requirements of each node.

In \Cref{sec:decrease-r} we formally describe the procedure. 
Given a proof labeling scheme $(f, \A)$ with radius $r$ certifying some property $\mathcal{P}$, we show how to construct a proof labeling scheme with radius $(r - \delta)$ certifying $\mathcal{P}$ with size bounded by $\Overhead$, where $\Delta$ is the maximum degree of the certified graph, $s$ is the certificate size of $(f, \A)$, and $\ell$ is the size of labels on the vertices of the input graph.
In \Cref{sec:lowerbound}, we show that the increase by a factor of $C \cdot \Delta^{\delta - 1}$ is necessary for some graph properties, while in Section~\ref{sec:shaving-logs} we prove that in some cases the multiplicative logarithmic terms can be replaced by an additive one.




\section{Preliminaries}
All the graphs are assumed to be undirected and simple with possible labels.
We also assume that all graphs are connected, as two different connected components have no way to interact with each other.
Formally $G = (V, E, L)$ where $L \colon V \to \bitstrings$.
The vertices are assumed to have assigned integer \emph{identifiers}, formally for a graph on $n$ vertices, we assume that $V = \{1, \dots, n\}$.

Neighbours of a vertex $v$ are denoted as $N_G(v)$, if $G$ is clear from the context, $N(v)$ is used. Distance between two vertices $u, v$ is denoted as $d_G(u, v)$, and the subscript is omitted if $G$ is clear from the context.
We denote the set of vertices within distance $r$ from $v$ as $V[v,r]$, also called the $r$-local neighbourhood of $v$.

A \emph{graph property} is formally a set of graphs that is closed under isomorphism, that is, its membership does not depend on the choice of identifiers.
Note that it may depend on the labels of the graph.
A \emph{certificate assignment} $P$ for $G$ is a function $P\colon V(G) \rightarrow \bitstrings$ that associates with each vertex a \emph{certificate}. 
We say that $P$ has size $s$ if $|P(v)| \leq s(n)$ for every $v$.
A \emph{verifier} is a function that takes as an input a graph $G$, its certificate assignment $P$ and $v \in V(G)$ and outputs either $0$ or $1$.
In this case, we say that the verifier is invoked on $v$.

We denote the induced subgraph $G[V[v,r]]$ as $G[v,r]$ and the restriction of $P$ to $V[v,r]$ is denoted as $P[v,r]$, that is $P[v,r] \colon V[v,r]\to \{0,1\}^*$. 
A verifier $\mathcal{A}$ is $r$-\emph{local} if $\mathcal{A}(G,P,v) = \mathcal{A}(G[v,r],P[v,r],v)$ for all $G$, $P$, and $v$. 


%


An $r$-local \emph{proof labeling scheme} certifying a property of labeled graphs $\mathcal{P}$ is a pair $(f,\mathcal{A})$,
where $\mathcal{A}$ is an $r$-local verifier and $f$, called the \emph{prover}, assigns to each $G \in \mathcal{P}$ a certificate assignment $P$ such that the following properties hold.
\begin{itemize}
    \item \emph{Completeness}: If $G \in \mathcal{P}$, then $\mathcal{A}(G[v, r], P[v, r], v) = 1$ for all $v$, where $P~=~f(G)$.
    \item \emph{Soundness}: If $G \notin \mathcal{P}$, then for every certificate assignment $P'$, there is $v$ such that $\mathcal{A}(G[v, r], P'[v, r], v)~=~0$.
\end{itemize}
We say that $(f, \mathcal{A})$ has size $s : \mathbb{N} \to \mathbb{N}$ if $|f(G)(v)| \leq s(|V(G)|)$ for all $G \in \mathcal{P}$ and all $v \in V(G)$.


\section{Decreasing the radius of a proof labeling scheme}\label{sec:decrease-r}

The goal of this section is to show that given an $r$-local proof labeling scheme $(f_r, \A_r)$ certifying property $\Prop$, we can construct an $(r-\delta)$-local proof labeling scheme $(f, \A)$ certifying $\Prop$ for any $\delta < r$ at the cost of increasing the certificate size by a certain amount.
The increase of the certificate size can be expressed as a function of the size of the input graph and its maximum degree.
The result is precisely formulated as follows.

\begin{theorem}\label{thm:decrease-r}
    Given an $r$-local proof labeling scheme $(f_r, \A_r)$ of size $s$ certifying a graph property $\mathcal{P}$, for every $\delta < r$, we can construct an $(r - \delta)$-local proof labeling scheme certifying $\mathcal{P}$ with certificates of size \[\Overhead\] where $\ell(n)$ is the maximum size of a label and $\Delta \geq 3$ is the maximum degree of the input graph.
\end{theorem}

Note that in the case of $\Delta = 2$, the maximum size of a $\delta$-neighborhood of a vertex grows only linearly with $\delta$ and we may obtain the bound on certificate size of $\mathcal{O}(\delta (\Delta \log(n) + s(n) + \ell(n)))$.

\subsection{Overview of the proof technique}
When the verifier $\A_r$ is invoked on $v$, it is given $G[v, r]$ and $P[v, r]$ on its input.
If we want to reduce that information to $G[v, r - \delta], P[v, r - \delta]$, a first step can be to \emph{move} the now missing information into the certificates.
The first obstacle comes from the fact that information in the certificates may not be true (as opposed to $G[v, r]$ provided on the input) and must be verified.



\subsection{Encoding neighborhood into certificates}

The essential idea is simple, we have each vertex hold its distance $\delta$ neighborhood in its certificate.
This allows other vertices within distance $r - \delta$ to gain information about the entire distance $r$ neighborhood and feed this information to the original $r$-local verifier.

Instead of having each vertex explicitly encode its $\delta$ neighborbood, we define the notion of \emph{packets}, which are then \emph{broadcast} into the extended neighborhood.
The effect is the same as encoding the $\delta$ neighborhood in the certificate, and the notion allows us to break down the verification into simple checks and the proof of correctness into simple observations.

\begin{definition}
We say that $\PacExp$ is a \emph{packet}, if
\begin{itemize}
\item $D$ is a set of vertices,
\item $L, C \in \bitstrings$,
\item $\omega$ is the identifier of so-called \emph{origin vertex},
\item and $d \in \{0, \dots, \delta\}$.
\end{itemize}
\end{definition}
Note that a packet can be easily encoded and decoded into a binary string.
Furthermore, it is easy to locally check that the individual elements of the packet correspond to the definition and have the correct types.

Intuitively, $D, L$, and $C$ are used to carry the local information around $\omega$.
In particular, $D$ will be the identifiers of $N(v)$, $L$ will be the label on $v$, and $C$ is a certificate on $v$ eventually passed on to $\A_r$.
We use $d$ to keep the distance of the packet from $\omega$ to ensure the correct distribution of the packet among other vertices.

The origin of a packet $p$ is denoted as $\omega(p)$, a similar notation is used with the other components of the packet.
Given a certificate assignment $P \colon V(G) \to \bitstrings $ and $v \in V$, $\Spacket{P, v}$ denotes the set of packets encoded in $P(v)$ (if $P(v)$ is not a valid encoding of packets then $\Spacket{P, v}~=~\emptyset$).
We define $\has{P, v, x}$ as true if and only if there is $p \in \Spacket{P, v}$ such that $\omega(p) = x$ and if true, we denote such packet $p$ as $\packet{P, v, x}$.

\begin{definition}
We say that packet $p = \PacExp$ is \emph{well-formed} if and only if $D = N_G(\omega)$ and $L$ is the label on $\omega$.
\end{definition}

Now, we show how well-formed packets can be used to reconstruct the neighborhood.
Let $\mathcal{B}$ be a set of packets and $\Orig{\mathcal{B}} \coloneqq \{ \Orig{p} \mid p \in \mathcal{B} \}$.
Let $G(\mathcal{B}) \coloneqq (\Orig{\mathcal{B}}, E(\mathcal{B}), L(\mathcal{B}))$, where 
$E(\mathcal{B}) \coloneqq \{ \{x, y\} \mid x, y \in \Orig{\mathcal{B}} \text{ and } \exists p \in \mathcal{B}$ such that $\Orig{p} = x \text{ and } y \in D(p) \}$
and $L(\mathcal{B})(v) = L(p)$ such that $\omega(p) = v$.

\begin{observation}\label{obs:reconstruct-from-packets}
    Let $\mathcal{B}$ be a set of well-formed packets such that $\Orig{\mathcal{B}} = \Nr{v}{r}$.
    Then $G(\mathcal{B}) = G[v, r]$.
\end{observation}
\begin{proof}
    For each $\{x, y\} \in E(G[v, r])$, there is $p \in \mathcal{B}$ with $\Orig{p} = x$ and $y \in D(p) = N(x)$ as $p$ is well-formed.
    On the other hand, for each $p$ with $\Orig{p} = x$ and $y \in D(p)$, there is $\{x, y\} \in E(G[v, r])$ if also $y \in V[v, r]$.
\end{proof}

Similarly, we can reconstruct the encoded certificate assignment, we define $\mathcal{C}(\mathcal{B}) : \Orig{\mathcal{B}} \rightarrow \bitstrings$ so that $C(\mathcal{B})(v) = C(p)$ where $p \in \mathcal{B}$ such that $\Orig{p} = v$.

\subsection{Constructing the ($r-\delta$)-local verifier}
Recall that we are given an $r$-local proof labeling scheme $(f_r, \A_r)$ certifying the property of labeled graphs $\mathcal{P}$.
We now define the $(r - \delta)$-local verifier.
The main task of the verifier is to ensure that all packets are distributed as needed and all are well-formed, that is they can be used to reconstruct wider neighbourhoods.

Let $\mathcal{B}[P,u, r - \delta] := \bigcup_{v \in V[u, r-\delta]} \Spacket{P, v}$, that is $\mathcal{B}[P, u, r - \delta]$ is the set of all packets that are visible from $u$ to distance $r - \delta$ and are encoded in $P$.
We define $\A$ so that $\mathcal{A}(G[v, r - \delta], P[v, r - \delta], v) = 1$ if and only if the following conditions are all satisfied:
\begin{enumerate}[label=\textbf{Condition B\arabic*:}, ref=Condition B\arabic*, leftmargin=*]
    \item \label{cond:packet-just-once} For each $x$, there is at most one $p \in \Spacket{P, v}$ such that $\Orig{p}~=~x$,
    \item \label{cond:packet-start} $(N(v), L(v), C', v, 0) \in \Spacket{P, v}$ for some $C' \in \bitstrings$,
    \item \label{cond:min-dist} if $\has{P, v, u}$ and $v \neq u$ then $1 \leq d(\packet{P, v, u}) = 1 + \min \{ d(\packet{P, x, u}) \mid \has{P, x, u} \text{ and } x \in N(v) \}$, 
    
    \item \label{cond:packet-propagate} $\has{P, v, u}$ for $v \neq u$ if and only if there is $x \in N(v)$ and $p' = \packet{P, x, u}$ such that $d(p') < \delta$,

    
    \item \label{cond:packet-consistent} for every $p = \packet{P, v, u} \in \Spacket{P, v}$ and every existing $p' = \packet{P, x, u}$ such that $x \in N(v)$, it holds $D(p) = D(p'), L(p) = L(p'), C(p) = C(p')$,
    \item \label{cond:orig-verifier-yes} let $G' = G(\visPackets{P, v, r - \delta})$ and $P' = C(\visPackets{P, v, r - \delta})$, then \\
    $\A_r(G'[v, r], P'[v, r], v) = 1$.
\end{enumerate}

Note that each condition requires information only about neighbours within distance at most $r - \delta$ and hence can be locally verified.
\ref{cond:packet-just-once} makes the reasoning easier, as we can assume that at most one packet from an originating vertex is present in the certificate.
\ref{cond:packet-start} ensures that every vertex has a well-formed packet originating from the vertex itself.
\ref{cond:min-dist} allows us to inductively prove that each packet correctly holds its distance from its originating vertex.
\ref{cond:packet-propagate} ensures that a packet originating from a vertex is distributed to vertices within a distance $\delta$ from it.
\ref{cond:packet-consistent} ensures that the packet is correctly \emph{copied} from one vertex to another.
\ref{cond:orig-verifier-yes} ensures that the original verifier accepts the described graph and certificate assignment. 

Now, we can establish properties of the encoded packets, given that the verifier accepts them.
To make the notation shorter, we denote $\mathcal{A}(G[v, r - \delta], P[v, r -\delta], v)$ as $\A(v)$.
First, we establish that the encoded distances are equal to the actual distances in the graph.

The following lemmas follow from the definitions of the six Conditions.
The reasoning is straightforward, so we have decided to move the proofs into an appendix. 
They are provided in \Cref{app:omittedProofs}.

\begin{lemma}\label{lem:local-to-global}
  If $\Adefault = 1$ for all $v$, then $d(\packet{P, u, v}) = d_G(u, v)$ for all $u, v \in V$ such that $\has{P, u, v}$.
\end{lemma}

Next, we establish that the local information of each vertex is indeed distributed into its $\delta$ neighborhood, given that the verifier accepts the certificates.
\begin{lemma}\label{lem:has-packets-from-delta}
    If $\Adefault = 1$ for all $v$, then for every $u, x \in V$ it holds $\has{P, u, x}$ if and only if $u \in V[x, \delta]$.
\end{lemma}

We now show that all packets with the same origin must hold the same local information.

\begin{lemma}\label{lem:cert-in-packets-same}
    If $\Adefault = 1$ for all $v$, then $C(\packet{P, v, x}) = C(\packet{P, u, x})$ for all $v, u, x$ such that $\has{P, v, x}$ and $\has{P, u, x}$.
\end{lemma}

Next, we show that the information carried by the packets must be well-formed.
\begin{lemma}\label{lem:packets-correct}
  If $\Adefault = 1$ for all $v$, then $\packet{P, v, x}$ is well-formed for every $v, x$ such that $\has{P, v, x}$.
\end{lemma}

It remains to establish that each vertex has access to enough packets to compute its distance $r$ neighborhood.

\begin{lemma}\label{lem:sees-all-packets-from-r}
    If $\Adefault = 1$ for all $v$, then for every $u \in V$ it holds $\Nr{u}{r}~\subseteq \Orig{\mathcal{B}[P, u, r - \delta]}$.
\end{lemma}

Next, we establish that if the original $r$-local verifier $\A_r$ accepts based on both the local information and the information in packets, the graph indeed satisfies the property $\Prop$.

\begin{lemma}\label{lem:proof-schema-sound}
    If $\Adefault = 1$ for all $v$, then $G \in \mathcal{P}$.
\end{lemma}

Finally, we are ready to prove the main theorem.
\begin{proof}[\Cref{thm:decrease-r}]
    It follows from \Cref{lem:proof-schema-sound} that $\A$ accepts $G$ only if $G \in \mathcal{P}$.
    It remains to show that any $G \in \mathcal{P}$ has a certificate assignment accepted by $\A$, with each certificate of size at most $\Overhead$.
    
    Given $G \in \mathcal{P}$, we construct $P$ so that for each $u$, $P(u)$ is an encoding of packets with one packet for each vertex $x \in \Nr{u}{\delta}$.
    A packet $p_x = \PacExp$ for vertex $u$ is constructed by setting $D \coloneqq N_G(x)$, $L \coloneqq L_G(x)$, $C \coloneqq f_r(G)(x)$, $\omega \coloneqq x$, and $d \coloneqq d_G(x, u)$.
    It remains to check that the verifier returns 1 on each vertex.

    Note that Conditions B1, B2, B3, B4, and B5 hold from the construction.
    Finally, note that by \Cref{obs:reconstruct-from-packets}, we have \[
    \A_r(G(\visPackets{P, v, r - \delta})[v, r], C(\visPackets{P, v, r - \delta})[v, r], v) = \A_r(G[v, r], f_r(G)[v, r]) = 1\]
    for each $v$ and hence \ref{cond:orig-verifier-yes} holds.
    Thus, the certificate assignment $P$ is accepted by $\A$.

    Now, we proceed to bound the maximum size of a certificate assigned by $P$.
    Assuming we encode a set of vertices as a set of their identifiers, each of size $\bigO(\log(n))$, the size of encoding of $\PacExp$ is bounded by
    $\bigO(\Delta(n) \log(n) + \ell(n) + s(n))$.
    We have $|\Spacket{P, v}| = |\Nr{v}{\delta}| \leq (\Delta (\Delta - 1)^\delta - 2) / (\Delta - 2) $ for $\Delta \geq 3$ and for all $v$, thus the size of encoding of $P(v)$ is bounded by $\Overhead$.
    
    In the case of $\Delta = 2$, we have $|\Nr{v}{\delta}| = 2 \delta + 1$ and then the size of encoding of $P(v)$ if bounded by $\mathcal{O}(\delta (\Delta \log(n) + s(n) + \ell(n)))$.
\end{proof}

\section{Lower bound on the increase of certificate size}\label{sec:lowerbound}


This section aims to show that there are proof labeling schemes for which the radius can be decreased by $\delta$ only if we also increase the certificate size by $C (\Delta-1)^{\delta  - 1}$, where $C$ is a fixed constant.
We present a property of labeled graphs, for which we also provide a proof labeling scheme and both an upper and a lower bound on its size.
Later, we describe a property of graphs without labels utilizing very similar idea.

\begin{figure}[t]
  \centering
  \includegraphics[width=0.8\textwidth]{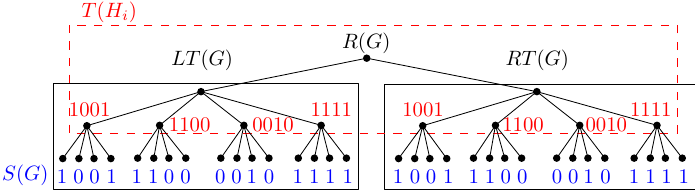}
  \caption{An example of a graph with property $\LbProp$ with $\Delta = 5$. Here, $R(G)$ is the root, $LT(G)$ and $RT(G)$ are the left and the right subtrees, $S(G)$ is the binary sequence in the leaves, and the red strings are certificates. The subgraph $T(H_i)$ is used in the proof of \Cref{lemma:lowerBound} and corresponds to $r = 2$. }
  \label{fig:example}
\end{figure}

Let $\Delta \geq 3$, then we define $\mathcal{P}_\Delta$ so that a labeled $G \in \mathcal{P}_\Delta$ if and only if it satisfies all of the following three properties.
For an example of a graph with the property, see \Cref{fig:example}.
\begin{property}[Tree structure]\label{prop:tree}
    $G$ has a single vertex of degree $2$, denoted as $R(G)$ (or just $R$, when a concrete $G$ is irrelevant or clear from the context), which is adjacent to two complete $(\Delta-1)$-nary trees of the same size.
\end{property}
\begin{property}[Label structure]\label{prop:labels}
    For every vertex $v$ except for the root $R(G)$, the label $L(v)$ encodes an integer $a \in \{1, 2, \dots, \Delta - 1\}$ that uniquely defines its order among its siblings.
    Additionally, if $deg(v) = 1$, then $L(v)$ also encodes one bit $b \in \{0, 1\}$.
    Therefore, on leaves $L(v)$ encodes a pair $(a, b)$.
    For $L(R(G))$, the label is empty.
\end{property}
This allows us to naturally define the \emph{left} and \emph{right} subtrees of $G$, i.e. the subtrees rooted on the first and second child of $R(G)$ respectively.
We denote those as $LT(G)$ and $RT(G)$.
Furthermore, it allows us to order the leaves of $G$.
We denote as $S(v)$ the binary string created by taking the values of $b$ on all leaves in their natural order in the subtree rooted at $v$.
We define $S(G) = S(R(G))$.
\begin{property}[String structure]\label{prop:string}
    $S(G) = XX$ for some binary string $X$, i.e. $S(G)$ is a result of concatenating a string $X$ with itself once.
\end{property}

Now we describe a proof labeling scheme of $\LbProp$.
The following lemma provides an upper bound on the optimal certificate size for a given radius. This is then used together with a corresponding lower bound, to show a lower bound on the necessary increase of the certificate size of $\mathcal{P}_{\Delta}$ when decreasing the radius.

\begin{lemma}\label{lem:lb-prop-up}
    Graph property $\mathcal{P}_\Delta$ has an $r$-local proof labeling scheme of size $\UBval$ for every $r \geq 1$ and a fixed $C$.
\end{lemma}
\begin{proof}
    We will show how to certify the Properties \ref{prop:tree}, \ref{prop:labels}, and \ref{prop:string}. 
    To certify Property \ref{prop:tree}, each vertex can be given in its certificates the identifier of the root (the vertex of degree 2), its distance from the root, and the depth of the tree.
    It is well known that distances are enough to certify the acyclicity of a graph and described for instance in \cite{feuilloley2021introduction}.
    To verify the depth of the whole tree, it suffices to ensure that each certificate has the same value of this depth and on leaves, this depth is verified to be equal to the distance from the root.

    To certify Property \ref{prop:labels}, it suffices to check the structure of each label.
    Checking the uniqueness of $a$ among siblings can be verified by each parent.


    To certify Property \ref{prop:string}, each vertex $v$ holds $S(v)$ in its certificate.
    This is again trivial to verify on a leaf and for any other $v$, assuming that all children $c_i$ hold correct $S(c_i)$, it suffices to check that $S(v)$ is their concatenation.

    Now the key idea of reducing the size of certificates is to encode the value of $S(v)$ only in the vertices with $d(v, R) \geq r$.
    Certificates of the vertices with $d(v,R) < r$ are left empty.
    Note that the number of leaves of the subtree of $v$ with $d(v, R) = k$ is at most $n / (2 (\Delta - 1)^{k - 1})$ as there are $(2 (\Delta - 1)^{k - 1})$ vertices with distance $k$ from $R$.
    Hence, we can encode $S(v)$ using at most $\mathcal{O}(n) / (\Delta - 1)^{k-1}$ bits.
    Now, $R$ may compute $S(G)$ as the concatenation of $S(v)$ for all $v$ with $d(v, R) = r$ and check that it satisfies \Cref{prop:string}.

    The rest of the properties require only $\mathcal{O}(\log(n))$ bits in certificates, thus there is an $r$-local proof labeling scheme that certifies $\LbProp$ with certificates of size at most $\UBval$ for some fixed $C$ and large enough $n$.
\end{proof}

Now, we show a lower bound on the required certificate size to locally certify~$\LbProp$.

\begin{lemma}\label{lemma:lowerBound}
    For all $r$-local proof labeling schemes certifying $\mathcal{P}_\Delta$ of size $s$, it holds \[
    s(n) \geq \LBvalBig
    \] for a large enough $n$ and all $\varepsilon < 1$.
\end{lemma}


\begin{proof}
    The idea of the proof is inspired by the proof of Theorem 6.1 of Göös and Suomela \cite{goos2016locally}.
    Following their approach, we will show that for every supposed proof labeling scheme of size less than $\LBval$, we can construct an instance not in $\LbProp$ which the verifier would necessarily accept.
    
    Suppose there exists an $r$-local proof labeling scheme $(\mathcal{A},f)$ certifying $\LbProp$ such that for every $n'$ there exists $n \geq n'$ such that $s(n) < \LBval$.
    For an instance $H_i \in \LbProp$, let $T(H_i)$ denote $V[R(H_i), r]$.
    Let $H_j \in \LbProp$ and let $\sim$ be a binary relation on $\LbProp$ defined so that $H_i\sim H_j$ if and only if $f(H_i)[T(H_i)] = f(H_j)[T(H_j)]$ and $H_i[T(H_i)] = H_j[T(H_j)]$,
    that is both the subgraphs on $T(H_i)$, $T(H_j)$, and their certificates as assigned by $f$ are the same.
    The equality of induced subgraphs here means the equality of the identifiers, the labels, and the edges.
    Note that $\sim$ is an equivalence.
    See again~\Cref{fig:example} for an illustration.
    
    Let $\LbProp[n]$ be the set of instances in $\LbProp$ on $n$ vertices and with a fixed identifier assignment, meaning the identifier of a vertex with a given position in the tree is the same in all the instances.

\begin{claim}
    For all $n'$, there exists $n \geq n'$ and $H_1, H_2 \in \LbProp[n]$ such that $H_1 \sim H_2$ and  $S(H_1) \neq S(H_2)$.
\end{claim}
\begin{proof}
    We will show that for large enough $n$, the number of possible binary sequences in the leaves of instances in $\LbProp[n]$ is greater than the number of equivalence classes of $\sim$ when restricted to $\LbProp[n]$.

    By the assumption, each vertex has less than $\LBval$ certificate bits, thus for an instance $H_i \in \LbProp[n]$, there are at most $2^{\LBval \cdot |T(H_i)|}$ different certificate assignments on $T(H_i)$, and at most $(\Delta-1)^{|T(H_i)|}$ different assignments of labels on $T(H_i)$.
    The rest of the structure on $T(H_i)$, including the identifiers is fixed by the fact that $H_i \in \LbProp[n]$.
    
    Furthermore, observe that $|T(H_i)| = 1 + 2\sum_{i=0}^{r-1} (\Delta-1)^i \leq 3(\Delta-1)^{r}$ as $\Delta \geq 3$.
    In total, we have that $\sim$ has on $\LbProp[n]$ at most $2^{(n \cdot \varepsilon) / 4 } \cdot (\Delta-1)^{3(\Delta-1)^r}$ different equivalence classes.

    On the other hand, each instance has at least $n/4$ leaves in the left subtree and thus there are at least $2^{n/4}$ different possible binary strings in the left subtree.
    It remains to observe that
    \[
    2^{(n\cdot\varepsilon) / 4  } \cdot (\Delta-1)^{3(\Delta-1)^r} < 2^{n/4}
    \]
    for large enough $n$.
    Therefore by the pigeonhole principle, there are $H_1, H_2 \in \LbProp[n]$ such that $S(H_1) \neq S(H_2)$ and $H_1 \sim H_2$.
\end{proof}

Now, we take $H_1, H_2 \in \LbProp[n]$ such that $H_1 \sim H_2$ and $S(H_1) \neq S(H_2)$ and construct $H' = (V', E', L')$ by starting with $H_1[T(H_1)] = H_2[T(H_2)]$ and completing the left subtree by $LT(H_1)$ and the right subtree by $RT(H_2)$.
Formally, let $L_S(G)$ be the neighbour of $R(G)$ in $LT(G)$ and $R_S(G)$ the neighbour in $RT(G)$. Then
\begin{align*}
V' =~& V(LT(H_1)) \cup V(RT(H_2)) \cup \{R(H_1)\} \\
E' =~& E(LT(H_1))\cup E(RT(H_2)) \cup \{R(H'),L_S(H_1)\} \cup \{R(H'),R_S(H_2)\}).
\end{align*} 
Observe that the identifier assignment of $H'$ is the same as those of $H_1$ and $H_2$, hence by construction, we have that $H'$ satisfies Properties \ref{prop:tree} and \ref{prop:labels} and the verifier can not reject $H'$ on their basis.
Furthermore, observe that $H' \notin \LbProp$ as the string in the leaves does not satisfy \Cref{prop:string}.

Now, we choose the certificate assignment on $H'$ as
\begin{equation*}
    P(v)=
    \begin{cases}
      f(H_1)(v) & \text{ if } v\in LT(H') \cup \{R(H')\} \\
      f(H_2)(v) & \text{ otherwise }
    \end{cases}
\end{equation*}
Recall that $f$ is the prover of the proof labeling scheme of $\LbProp$.


\begin{claim}
    For all $v\in V(H')$ it holds $\mathcal{A}[H'[v,r],P[v,r]]=1$.
\end{claim}
\begin{proof}
    First, recall that $T(H') = T(H_1) = T(H_2)$, and by construction of $P$, we also have $P[T(H')] = f(H_1)[T(H_1)] = f(H_2)[T(H_2)]$.
    Observe that if $v \in LT(H') \cup \{R(H')\}$, we have $H'[v,r] = H_1[v,r]$ and $P[v,r] = f(H_1)[v,r]$ and thus $\mathcal{A}[H'[v,r],P[v,r]] = 1$.
    Similarly, if $v \in RT(H')$, we have $H'[v,r] = H_2[v,r]$ and $P[v,r] = f(H_2)[v,r]$ and thus $\mathcal{A}[H'[v,r],P[v,r]] = 1$.
\end{proof}
Hence, there is an instance $H' \notin \LbProp$ which is accepted by $\mathcal{A}$, contradicting the assumption that $(f, \mathcal{A})$ certifies $\LbProp$.
This finishes the proof.
\end{proof}




Similar approach can be used in a graph property without labels.
The idea behind it is to encode the labels using some subgraphs.

We denote the optimal certificate size of an $r$-local proof labeling scheme for a property $\mathcal{P}$ as $s^*(\mathcal{P}, r, n)$.
An $r$-local proof labeling scheme has optimal certificate size $s$ if there is no other $r$-local proof labeling scheme of size $s'$ such that there is $n$ with $s'(n) < s(n)$.


\begin{figure}[t]
  \centering
  \includegraphics[width=0.8\textwidth]{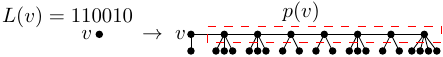}
  \caption{An example of a vertex $v$ in a graph $G\in \mathcal{P}^l$ and the same vertex in $g(G)$}
  \label{fig:label2path}
\end{figure}

\begin{lemma}\label{lem:no-need-for-labels}
    For every property $\mathcal{P}^l$ with $r$-local proof labeling scheme of at most polynomial and at least logarithmic size, there exists a property $\mathcal{P}^u$ with no labels, except for unique identifiers, such that
    \[
        s^*(\mathcal{P}^l, r, n) \leq c_1 s^*(\mathcal{P}^u, r, n) \leq c_2 s^*(\mathcal{P}^l, r, n),
    \]
    for every $r$ and every large enough $n$, and where $c_1, c_2$ are fixed positive constants.
\end{lemma}

The result is not surprising and the idea of the proof is simple but technical, therefore we decided to move the proof to \Cref{sec:prooflemma}.
An example of the encoding of a label into a subgraph can be seen in the \Cref{fig:label2path}.
A single leaf vertex and a path $p(v)$ is attached to each vertex $v$.
The $i$-th vertex of $p(v)$ has $2$ or $3$ leaf neighbours depending on the $i$-th bit of $L(v)$.
The last vertex represents the end of the string. 
We then assume that we are provided with a proof labeling scheme for the unlabeled property and make the decision based on this scheme.

Now, we are ready to prove there are proof labeling schemes, such that the increase of certificate size by $C (\Delta - 1)^{\delta - 1}$ is necessary when decreasing the radius by $\delta$.


\begin{theorem}\label{thm:lowerUpper}
    There is an $r$-local proof labeling scheme, certifying a property without labels, of size $s_r$ such that after decreasing its radius by $\delta$, it holds for any possible resulting $r - \delta$-local proof labeling schema of size $s_{r - \delta}$ and every large enough $n$
    \[s_{r - \delta}(n) \geq s_r(n) \cdot C (\Delta-1)^{\delta  - 1}\]
    where $\Delta$ is the maximum degree of the input graph and $C$ is a fixed constant.
\end{theorem}
\begin{proof}
    Consider the property $\LbProp$.
    By \Cref{lem:lb-prop-up}, it can be certified by an $r$-local proof labeling schema of size $s_r$ with $s_r(n) \leq C' \cdot n / (\Delta-1)^{r - 1}$ for every large enough $n$.
    By \Cref{lemma:lowerBound}, we have
    \begin{align*}
        s_{r - \delta}(n) \geq &~(n\cdot \varepsilon) / (12 (\Delta - 1)^{r - \delta} ) \geq \frac{ \varepsilon}{12 C'} \cdot s_r(n) \cdot  (\Delta-1)^{r - 1 - (r - \delta)}  = \\
        = &~s_r(n) \cdot C (\Delta-1)^{\delta - 1}
    \end{align*}
    for every large enough $n$ and a fixed $C$.

    By \Cref{lem:no-need-for-labels}, for any $r$ and $\mathcal{P}_\Delta$, there exists a property $\mathcal{P}^u$ with optimal certificate size $s^u_r$ of an $r$-local proof labeling scheme such that:
     $C_1 s^u_{r}(n) \geq s_{r}(n) \geq C_2 s^u_r(n)$.
     Applied on both $s_r$ and $s_{r-\delta}$, we obtain
     \[
     C_1 s^u_{r - \delta}(n) \geq s_{r - \delta}(n) \geq s_r(n) C(\Delta - 1)^{\delta - 1} \geq s^u_r(n) C_2 \cdot C(\Delta - 1)^{\delta - 1}.
     \]

\end{proof}

\section{Saving on the log factors on paths}
\label{sec:shaving-logs}

In Section~\ref{sec:lowerbound}, we proved that in general when reducing the verification radius, a blow-up of the certificate size is necessary. Roughly, the size of the certificates for radius 1 needs to be multiplied by the size of the ball of radius $d$, compared with certificates for radius $d$.
In the expression of Theorem~\ref{thm:decrease-r}, there are two additional terms: one related to the input labeling (which seems difficult to remove), and one related to identifiers, of the form ``size of the ball'' multiplied by $O(\log n)$. 
In this section, we show that when the graph is a path, this last term can be replaced by a simple additive $O(\log n)$. In other words, it is not always necessary to encode locally all the identifiers of the ball at distance $d$.

\begin{theorem}\label{thm:shaving-logs}
Consider a property $\mathcal{P}$ on paths.
If there exists a proof-labeling scheme at distance $d$ for $\mathcal{P}$ using certificates of size $s(n)$, then there exists a proof-labeling scheme at distance 1, with certificates of size $O((2d+1)s(n)+\log n)$.
\end{theorem}

Note that without additional work, Theorem~\ref{thm:decrease-r}  gives size $O((2d+1)s(n)+(2d+1)\log n)$ (forgetting about input labels). 

The full proof is deferred to Appendix~\ref{app:shaving-logs}, we only give the intuition here. In the general approach to reduce the verification radius, we need to write the identifiers of all the nodes of the ball at distance $d$ in the new certificates (in addition to the old certificates and the inputs). This is because otherwise the new verifier would not be able to safely simulate the old verifier. Now, if we imagine that the identifier assignment is fixed, and not adversarial, we could go without that: a node would only check its own identifier and deduce safely the identifiers of its neighbors (forgetting about symmetry issues). What we prove is that if the graph is a path, then the prover can give and certify a new identifier assignment, 1, 2, 3, \dots, and give certificates related to this virtual assignment. And then the verifier can check that it would accept in the virtual identifier assignment, which is enough to prove the correctness of the input graph. This requires the prover to give only one identifier in the new certificates (the one of the node at hand) and not all the identifiers of the ball, which makes the multiplicative $O(\log n)$ become an additive $O(\log n)$. We refer the reader to the appendix for discussion of this ``virtual identifier'' technique, which, as far as we know, has never been used.


\section{Conclusion}
There are several remaining interesting open questions regarding the role of radius in local certification.
 An open question is the price of reducing radius considering some other family of graphs than bounded degree graphs. 
 An example of such a family may be planar graphs (degeneracy).

Another question to consider is the price of decreasing radius depending on the properties being certified.
While our approach works in general, there may be more efficient certification methods for specific properties.

\medskip
\noindent\textbf{Acknowledgment.} This work was supported by the Grant Agency of the Czech Technical University in Prague, grant No. SGS23/205/OHK3/3T/18 and by the Czech Science Foundation Grant no. 24-12046S.




\bibliography{ref}

\appendix 
\section{Omitted proofs from \Cref{sec:decrease-r}}\label{app:omittedProofs}
In this appendix, proofs of several lemmas from \Cref{sec:decrease-r} are provided. 

\begin{proof}[\Cref{lem:local-to-global}]
    Assume the lemma hypothesis holds.
    Observe that by \ref{cond:packet-just-once} and \ref{cond:packet-start} we have that $d(\packet{P, u, v}) = 0$ if and only if $u = v$ and $d_G(u, v)~=~0$.
    From now on, let us use the notation $\packet{u} = \packet{P, u, v}$.

    Suppose there is $u \in V$ such that $d(\packet{u}) < d_G(v, u)$, let $u$ be such a vertex with minimum $d(\packet{u})$.
    By \ref{cond:min-dist} there is $x \in N(u)$ such that $d(\packet{x}) = d(\packet{u}) - 1$.
    Note that $u \neq v$.
    It follows from the choice of $u$ that $d(\packet{x}) \geq d_G(x, v)$ and thus $d(\packet{u}) < d_G(v, u) \leq d_G(x, v) + 1 \leq d(\packet{x}) + 1 = d(\packet{u})$, a contradiction.

    Now suppose there is $u \in V$ such that $d(\packet{u}) > d_G(v, u)$, let $u$ be such a vertex with minimum $d_G(v, u)$.
    As $u \neq v$, there is $y$ such that $d_G(u, v) = d_G(y, v) + 1$.
    It follows from the choice of $u$ that $d(\packet{y}) \leq d_G(y, v)$ and by the previous argument $d(\packet{y}) = d_G(y, v)$.
    Hence, $d_G(v, y) = d(\packet{y}) = d_G(v, u)~-~1 < d(\packet{u}) - 1$ and thus $d(\packet{y}) + 1 < d(\packet{u})$, which contradicts $\A(u) = 1$ due to \ref{cond:min-dist}.
\end{proof}

\begin{proof}[\Cref{lem:has-packets-from-delta}]
Let there be $x \in V$ and $u \in \Nr{x}{\delta}$ such that $\neg \has{P, u, x}$, assume that $u$ is such vertex with minimum $d_G(u, x)$.
    Note that $u \neq x$, otherwise $\A(u) = 0$ due to~\ref{cond:packet-start}.
    
    Therefore, there is $y \in N(u)$ such that $d_G(u, x) = d_G(y, x) + 1$ and thus $\has{P, y, x}$ by the choice of $u$, and by \Cref{lem:local-to-global} $\packet{P, y, x}$ has encoded the correct distance.
    Hence, we have $y \in N(u)$ such that $\has{P, y, x}$, $d_G(y, x) = d(\packet{P, y, x}) < \delta$, and $\neg \has{P, u, x}$, thus $\A(u) = 0$ due to \ref{cond:packet-propagate}, a contradiction.

    Now suppose there is $x \in V, u \notin \Nr{x}{\delta}$ such that $\has{P, u, x}$, let $u$ be such vertex with minimum $d(u, x)$.
    As $u \neq x$, it follows from \ref{cond:min-dist} that there is $y \in N(u)$ with $d(\packet{P, y, x}) = d(\packet{P, u, x}) - 1$.
    By the choice of $u$, we have $y \in \Nr{x}{\delta}$ and by \Cref{lem:local-to-global} $\packet{P, y, x}$ has encoded the correct distance.
    We have $\delta < d_G(u, x) \leq d_G(y, x) + 1 \leq \delta + 1$, and hence $d_G(u, x) = d_G(y, x) + 1$.
    Therefore $d(\packet{P, y, x}) = \delta$ and $\A(u) = 0$ due to \ref{cond:packet-propagate}, a contradiction.
\end{proof}

\begin{proof}[\Cref{lem:cert-in-packets-same}]

Suppose there are $v, u, x \in V$ such that $C(\packet{P, v, x}) \neq C(\packet{P, u, x})$.
    By \Cref{lem:has-packets-from-delta}, there is $\packet{P, y, x}$ if and only if $y \in \Nr{x}{\delta}$, and therefore there are two incident vertices $v', u' \in \Nr{x}{\delta}$ such that $C(\packet{P, v', x}) \neq C(\packet{P, u', x})$.
    Then $\A(v') = 0$ due to \ref{cond:packet-consistent}, a contradiction.
\end{proof}

\begin{proof}[\Cref{lem:packets-correct}]
    Suppose there is $p = \packet{P, v, x}$ that is not well-formed, that is $D(p) \neq N(x)$ or $L(p) \neq L(G)(x)$.
    Note that due to \ref{cond:packet-start}, there is $\packet{P, x, x}$ which is well-formed.
    Due to \Cref{lem:has-packets-from-delta} and the connectedness of $G[x, \delta]$, there are two incident $u', v' \in \Nr{x}{\delta}$ such that
    $\packet{P, u', x}$ is well-formed and $\packet{P, v', x}$ is not well-formed. 
    It follows that $\A(v') = 0$ due to \ref{cond:packet-consistent}, a contradiction.
\end{proof}

\begin{proof}[\Cref{lem:sees-all-packets-from-r}]
    It follows from $\Adefault = 1$ and \ref{cond:packet-start} that every vertex has a packet originating from itself, thus $\Nr{u}{r - \delta} \subseteq \Orig{\mathcal{B}[P,u, r - \delta]}$ for every $u$.
    For every $x \in V$ with $r - \delta < d(u, x) \leq r$, note that on a shortest path between $u$ and $x$, there is $y$ such that $d(u, y) + d(y, x) = d(u, x)$ and $d(u, y) = r - \delta$.
    It follows that $d(y, x) = d(u, x) - d(u, y) \leq \delta$.
    Hence by \Cref{lem:has-packets-from-delta}, it holds $\packet{P, y, x} \in \mathcal{B}[P, u, r - \delta]$ and thus $x \in \Orig{\mathcal{B}[P, u, r - \delta]}$.
\end{proof}

\begin{proof}[\Cref{lem:proof-schema-sound}]
    Suppose that $P$ is the certificate assignment so that $\Adefault = 1$ for all $v$, then we claim that there is $P'$ such that $\A_r(G[v, r], P'[v, r], v) = 1$ for all $v$ and thus $G \in \mathcal{P}$.
    Let $P'(u) = C(\packet{P, u, u})$ and note that by \Cref{lem:cert-in-packets-same}, we have that $C(\packet{P, v, u}) = P'(u)$ for all $v$ with $\has{P, v, u}$.
    Furthermore, by \ref{cond:packet-start} it holds $\has{P, u, u}$ for all $u$, hence $P'$ is well-defined.

    Now let $v \in V$, then by \Cref{lem:sees-all-packets-from-r} $v$ has access to a packet originating from each vertex of $V[v, r]$.
    Each such packet is well-formed by \Cref{lem:packets-correct}, therefore by \Cref{obs:reconstruct-from-packets} we have $G(\Spacket{P, v}) = G[v, r]$.
    Due to \ref{cond:orig-verifier-yes}, we have $\A_r(G(\Spacket{P, v}), C(\Spacket{P, v}), v) = \A_r(G[v, r], P'[v, r]) = 1$ and hence $G \in \mathcal{P}$.
\end{proof}

\section{Proof of \Cref{lem:no-need-for-labels}}\label{sec:prooflemma}

We will define a graph property of unlabeled graphs and show mappings between the properties. 
Throughout the following proof, we always consider a fixed radius.
When $\mathcal{P}$ and $r$ are clear from the context, we use simply $s^*(n)$ instead of $s^*(\mathcal{P}, r, n)$.

\begin{proof}
The goal to show that for any $P^l$, we can construct $P^u$ such that the optimal $r$-local proof labeling schemes of these two properties have certificate sizes that differ by a constant factor.

This is achieved in two steps: first, we construct a proof labeling scheme of $P^u$ which essentially simulates the optimal proof labeling scheme of $P^l$ with a small overhead in the size of certificates. Subsequently, we also construct a proof labeling scheme of $P^l$ which simulates the optimal proof schema of $P^u$, again with only a small overhead.

We now define $P^u$. For that end, we show how to transform each $G \in P^l$ and encode its labels into the graphs structure, the resulting graph will be denoted as $g(G)$.

We define $g(G)$ the graph constructed from $G$.
For an example see \Cref{fig:label2path}
First, to each vertex $v\in V$ a leaf vertex is attached.
Then, for each $v\in V$ a path $p(v)$ is created of the length of $|L(v)|+1$ and attached to $v$. 
Let $p(v)_i$ be the $i$-th vertex of the path $p(v)$ and $L(v)_i$ the $i$-th bit of the label $L(v)$. 
A certain number of leaf vertices is attached to each vertex of $p(v)$ in the following way:
\begin{itemize}
    \item if $L(v)_i = 0$ then $2$ leaves are attached to $p(v)_i$,
    \item if $L(v)_i = 1$ then $3$ leaves are attached to $p(v)_i$,
    \item if $i = |L(v)|+1$ then $4$ leaves are attached to $p(v)_i$.
\end{itemize}
Note that as the property $P^l$ is closed under permutation of identifiers, the resulting graph may also have any choice of identifiers.
For convenience, we assume that the identifiers of the \emph{original} vertices of $G$ are preserved in $g(G)$.
The set of graphs created by this procedure from $P^l$ is denoted as $P^u$ (with all possible identifier permutations).
By the $L(G)$ we denote the largest label on graph $G$.

\begin{claim}
    It holds that $|V(g(G))| \leq 5|V(G)| \cdot |L(G)+1|$.
\end{claim}
\begin{proof}
    For each vertex, one leaf is added and a path of the size of the labels with at most 4 additional leaves for each vertex of the path. 
    Together that is $1+4|L(G)+1|$ vertices per vertex of the original graph.
    Together that is $|V(G)| \cdot (1+4|L(G)+1|)$.
    For convenience we simplify the second factor to $5|L(G)+1|$.
\end{proof}

This increase in the number of vertices is the reason for the at most polynomial certificate size. 
If we considered certificates with larger size, then the unlabeled graph would be allowed to have asymptotically larger certificates. 
As any unlabeled property can be certified with certificates of size $O(n^2)$, this restriction is justified.

Now, we are ready to define an $r$-local proof labeling scheme $(f^u, A^u)$ of $P^u$ based on the optimal proof labeling scheme of $P^l$, say $(f^{l*}, A^{l*})$.
We shorten $s^*(\mathcal{P}^l, r, n)$ to $s^*_l(n)$, similarly with $s^*_u$.
Note that following arguments do not change with $r$ and thus hold for every $r$.
\begin{claim}
  $s^*_u(n) \leq c_2 s^*_l(n)$ for a fixed $c_2$.
\end{claim}
\begin{proof}

Let $H\in \mathcal{P}^u$ and $G\in \mathcal{P}^l$ be graphs such that $g(G) = H$. 
We define $g'(H)=G$.

The certificates assigned by $f^u$ are tuples $(b_l,b_p,s,o)$.
\begin{itemize}
    \item $b_l$ is one bit denoting whether a vertex $v$ is a leaf or not. 
        If $v$ is a leaf, the rest of the certificate is empty.
    \item $b_p$ is one bit denoting whether a vertex $v$ is a vertex from $p(w)$ for some vertex $w$.
    \item $s$ is a bit string from which the labels are constructed.    
    \item If $v$ has exactly one leaf neighbour, $o = f^{l*}(g'(H))(v)$.
\end{itemize}
The verifier $\mathcal{A}^u$:
\begin{itemize}
    \item On a leaf vertex it accepts if and only if $b_l = 1$.
    \item On a vertex with $4$ leaf neighbours it accepts if and only if $b_p = 1$ and $s$ is empty. 
    \item On a vertex with $3$ leaf neighbours accepts if and only if:
    \begin{itemize}
        \item it has two non-leaf neighbours, one, denoted as $w$, with the element $s$ one bit shorter, the other either with $b_p=0$ and the same $s$ or with $b_p=1$ and $s$ one bit longer.
        \item the element $s$ is a concatenation of a $1$ and the element $s$ of $w$.
    \end{itemize} 
    \item On a vertex with $2$ leaf neighbours accepts if and only if:
    \begin{itemize}
        \item it has two non-leaf neighbours, one, denoted as $w$, with the element $s$ one bit shorter, the other either with $b_p=0$ and the same $s$ or with $b_p=1$ and $s$ one bit longer.
        \item the element $s$ is a concatenation of a $0$ and the element $s$ of $w$.
    \end{itemize} 
    \item On a vertex with $1$ leaf neighbour accepts if and only if:
    \begin{itemize}
        \item $b_p = 0$,
        \item it has exactly one neighbour with $b_p=1$ and the element $s$ is the same as the element $s$ of that neighbour,
        \item $\mathcal{A}^l$ executed for each $v$ with labels corresponding to $s$ and certificates to $o$ accepts on the local neighbourhood $g'(H)[v,r]$.  
    \end{itemize}
\end{itemize}

Now we show the size of $(f^u, \mathcal{A}^u)$.
The elements $b_l$ and $b_p$ consist of one bit and $o$ is of the same size as the proof labeling scheme $(f^l, \mathcal{A}^l)$. 
Recall that in $\mathcal{P}^u$ the structure of $p(v)$ for every $v$ is of a constant size relative to the size of the whole graph. 
The element $s$ of a vertex $v$ consists of as many bits as there are vertices in $p(v)$.
Hence $s^*_u(n) \leq c_2 s^*_l(n)$ for a fixed $c_2$.

\end{proof}

\begin{claim}
  $s^*_l(n) \leq c s^*_u(n)$ for a fixed $c$.
\end{claim}
\begin{proof}

We now define a proof labeling scheme $(f^l, A^l)$ based on the optimal proof labeling scheme of $P^u$, say $(f^{u*}, A^{u*})$.
  We want to encode the neighborhood of each $v \in g(G)$, including its $p(v)$, leaves and certificates of all of these into the certificate of $v \in G$.

For each graph $G\in \mathcal{P}$ the certificate assignment $f^l(G)$ constructs the corresponding $g(G)$ and gives it as an input to $f^u$.
For each $v \in V$ the certificate $f^l(G)(v)$ consists of $f^u(g(G))(v)$ and an encoding of $p(v)$ including certificates of any vertex $w\in p(v)$, as well as the certificate of the one leaf neighbour of $v$. 

On each $v$, the verifier $A^l$ takes the certificate of $v$ and constructs from it the local subgraph $g(G)[v,r] \cup p(v)$ with the certificates and verifies that the path $p(v)$ and the leaves attached to it were constructed correctly from the label $L(v)$.
It then executes $A^u$ on each vertex of the reconstructed graph. 
If $\mathcal{A}^u$ would reject on any vertex, the verifier $\mathcal{A}^l$ rejects well. 


As mentioned above, the subgraph $p(v)$ and the leaves adjacent to any vertex in $p(v)$, along with the certificates and identifiers of those vertices, are encoded in the certificate of any vertex $v$. 
This gives us that the size of $f^l(G)(v)$ is at most $5 |L(v) + 1| \cdot (|f^u(g(G))| + \log(n))$.
The $\log(n)$ accounts for the identifiers of the encoded vertices. 

Thus we have proven that $\mathcal{A}^l$ accepts if and only if $\mathcal{A}^u$ accepts and by the first claim it holds $c_1s^*_l(n)\leq s^*_u(n)$ for some $c_1>0$ as $c_1s_l(n)\leq s_u^*(n)$, where $s(n)$ is the size of $f^l$.

\end{proof}

And hence the lemma is proven.

\end{proof}

\section{Proof of \Cref{thm:shaving-logs}}
\label{app:shaving-logs}

Let us first remind the statement. 

\medskip{}

\textbf{Theorem.}
Consider a property $\mathcal{P}$ on paths.
If there exists a proof-labeling scheme at distance $d$ for $\mathcal{P}$ using certificates of size $s(n)$, then there exists a proof-labeling scheme at distance 1, with certificates of size $O((2d+1)s(n)+\log n)$.

\medskip{}

The core of the proof of this statement is not specifically about reducing the checkability radius, but about working on a worst case identifier assignment versus working on a more structured identifier assignment. 
We first state and prove a lemma capturing this intuition, and then show how this adapts to our context.
 
Throughout this section, suppose the graph is a path. The two \emph{canonical identifier assignments} of a path, whose endpoints are denoted $a$ and $b$, are the ones where the identifier of every node $u$ is its distance to $a$ (resp. $b$). 
(In other words the identifiers are 1,2,3,... in the order of the path.) 
Let a \emph{weak local certification} be the same as a local certification, except that the identifier assignment is promised to be canonical. 

\begin{lemma}\label{lem:canonical-to-general}
Consider a property $\mathcal{P}$ on paths. Any weak local certification with certificates of size $s(n)$ and radius $r$ can be turned into a (standard) local certification with certificates of size $O(s(n)+\log n)$ and radius $r$. In addition, the new certification is independent of the identifier assignment.
\end{lemma}

\begin{proof}[Proof of Lemma~\ref{lem:canonical-to-general}]
Let us first describe the behavior of the new prover $f'$ on correct instances, based on the old prover $f$. The prover $f'$ first chooses one of the canonical identifier assignments, call it $I^*$. Then it assigns to every node~$v$ a first label containing $I^{*}(v)$. Finally, the new prover $f'$ simulates the old prover on $(G,I^*)$ (which is possible since $I^*$ is canonical), that is, it assigns a second label to every node $v$ which is $f(G,I^{*},v)$.\footnote{Since we are modifying the identifier assignment, we make it appear explicitly in the notation for this proof.}
The certificate of a vertex is the concatenation of the two labels.

We now describe the verifier. The new verifier $\mathcal{A'}$ first checks that the certificates it sees are composed of two parts: one encoding a new identifier, and another label (if it is not the case, it rejects). For a vertex $v$, let us call these two parts $J(v)$ and $L(v)$.
The verifier $\mathcal{A}'$ then checks that $J$ corresponds to a canonical ordering in its view (this is straightforward). 
Finally, it simulates the old verifier taking $J$ as the identifier assignment, and $L$ as certificates.  
More formally, if it has not rejected yet, then it outputs $\mathcal{A}(G,L,J,v)$.

Let us prove that this scheme is correct.
Suppose that $\mathcal{A}'$ accepts on all the vertices of a given path with a given certificate assignment. Then this certificate assignment can be decomposed into a proper canonical ID assignment, and a another labeling which makes $\mathcal{A}$ accept. Since $\mathcal{A}$ is a correct proof-labeling scheme when the ID assignment is canonical, this path must be correct.
Conversely, for a negative instance, either the certificates are not properly encoded, or the encoded ID assignment is not canonical, or $\mathcal{A}$ rejects, which leads to a reject in at least one vertex.

The new proof labeling scheme is independent of the identifier assignment, since these are never used in the scheme.
\end{proof}

We now adapt the scheme to prove our theorem.

\begin{proof}[Proof of Theorem~\ref{thm:shaving-logs}]
The proof of the theorem is a mix of the proof of the previous lemma and of the general intuition for reducing the radius.
We start with the scheme $(f,\mathcal{A})$ at distance $d$. First, we build the scheme $(f',\mathcal{A}')$ at distance $d$ given by the proof of Lemma~\ref{lem:canonical-to-general}. Second, we use the general technique described in the previous sections to reduce the verification radius: a new prover $f''$ gives to every vertex all the certificates that $f'$ would give to the vertices in the ball at distance $d$, along with the structure of the graph and the identifiers. Finally, to save on the certificate size, we can remove all identifiers from the certificates, except for the one of the vertex itself. 
In other words, a vertex receives in its certificate: its new identifier, the certificates that $f'$ would assign to its neighbours at distance at most $d$, and the structure of this neighbourhood.
The key point is that since the identifier assignment used in $(f',\mathcal{A}')$ is canonical, the vertices can easily check the consistency with their neighbours.
Then the verifier $\mathcal{A''}$ will simply use the canonical ordering and proceed similarly to the verifier of the proof of Theorem~\ref{thm:decrease-r}.
\end{proof}

We expect that similar techniques can be used for other very structured graphs, such as grids. To which extent it is possible to generalize such a theorem is a very intriguing question. It has a similar flavor as the trade-off conjecture in the case of colorings~\cite{BousquetFZ24}, in the sense that in both cases the question boils down to how well we can recover a node labeling in a graph if we are given only some of these labels (in the current proof the labels are the identifiers, while in ~\cite{BousquetFZ24} these are the colors). See also~\cite{BalliuBKNORS24} for similar questions.

\end{document}